\documentclass[dvips]{acta}
\usepackage{supertabular,lscape,epsfig}
\usepackage{amssymb}
\usepackage{amsmath}

\SetPages{0}{0}

\SetVol{59}{2009}

\begin{document}

\begin{Titlepage}

\Title{\bf Photometric Study of Variable Stars in the Open Cluster NGC\,6866}

\Author{J.~M~o~l~e~n~d~a~-~\.Z~a~k~o~w~i~c~z, G.~K~o~p~a~c~k~i, M.~S~t~\k{e}~\'s~l~i~c~k~i, and A.~N~a~r~w~i~d}
{Astronomical Institute, University of  Wroc{\l }aw, Kopernika 11, Wroc{\l }aw, Poland\\
e-mail: (molenda,kopacki)@astro.uni.wroc.pl}

\Received{Month Day, Year}
\end{Titlepage}

\Abstract{We report the discovery of 19 variable stars and two blue--stragglers in the field of 
the open cluster NGC\,6866.
Three of the variable stars we classify as $\delta$\,Sct, two, as $\gamma$ Dor, 
four, as W\,UMa, two, as ellipsoidal variables, and one, as an eclipsing binary. Seven stars 
show irregular variability. Two of the pulsators, a $\delta$~Sct star NGC\,6866-29
and a $\gamma$ Dor star NGC\,6866-21, are multiperiodic.

From an analysis of proper motions, we conclude that the $\delta$\,Sct stars, one of the $\gamma$ Dor 
stars and both blue--stragglers are very probable members of the cluster. The position 
on the color-magnitude diagram of seven other variables suggests that they also belong to the cluster. 
The eclipsing binary, 
which we discover to be a new high-velocity star, and the seven irregular variables are non-members.

Then, we discuss in detail the age and metallicity of open clusters that host $\gamma$ Dor stars and 
we show that none of these parameters is correlated with the number of $\gamma$ Dor stars in cluster.
}
{Stars: pulsating: $\delta$ Sct -- Stars: pulsating: $\gamma$ Dor -- Stars: variable: 
other -- Open clusters: individual: NGC\,6866 -- Space missions: Kepler}

\section{Introduction}

NGC\,6866 $(\alpha _{\rm 2000}=20^{\rm h}03^{\rm m}55^{\rm s}$, $\delta _{\rm 2000}=
44^\circ09^\prime30^{\prime\prime})$ is an intermediate-age open cluster classified as I\,2p 
by Trumpler (1928) or II\,2m by Ruprecht (1966). The cluster falls into the field of view of 
the Kepler satellite telescope which will observe selected stars from the Cygnus--Lyra region with 
the aim of searching for Earth-size planets and a detailed study of the structure of 
pulsating stars by means of the asteroseismic analysis (Christensen-Dalsgaard \etal 2006).

The first photometric study of the cluster dates back to the $UBV$ photoelectric and 
photographic photometry of Hoag \etal (1961), Johnson \etal (1961), and Barkhatova \& Zakharova (1970). 
Hoag \etal (1961) and Johnson \etal (1961) derived the distance modulus of the cluster, $V-M_V = 10.82$ mag, 
the color excess, $E(B-V)=0.14$ mag, the distance, $d = 1200$ pc,
and estimated the spectral type of the turn-off point to be A3.
Subsequent observations of Sutantyo \& Hidajat (1972) yielded $E(B-V)=0.16$ mag 
and $V-M_V = 11.10$ mag. The catalogue of astrophysical data for Galactic 
open clusters of Kharchenko \etal (2005) gives $E(B-V) = 0.17$ mag, $d = 1450$ pc, 
and $V-M_V = 11.33$ mag. 

The estimates of the age of NGC\,6866 range from 0.23 Gyr (Lindoff 1968) to
0.65 Gyr (Loktin \etal 1994) with the most recent determination, 0.48 Gyr, 
given by Kharchenko \etal (2005). 
The mean $\rm [Fe/H]$ of the cluster has been derived by Loktin \etal (1994)
from the photometry of Hoag \etal (1961) to be equal to $+0.10$ dex.
The mean radial-velocity of the cluster has been measured by Mermilliod \etal (2008) and by 
Frinchaboy \etal (2008); these authors found the radial-velocity to be equal to 
$13.68\pm0.06$ and $12.18\pm0.75$ km/s, respectively. (Here, the uncertainty given by 
Mermilliod \etal (2008) is an r.m.s.\ error, while the uncertainty given by Frinchaboy \etal (2008),
the standard deviation.)

The probability of the cluster membership has been computed by Baumgardt \etal (2000)
for two stars, by Dias \etal (2002), for 89 stars, by Kharchenko \etal (2004), for 192 stars, and
by Frinchaboy \etal (2008), for 52 stars (all in common with Dias \etal (2002)). 
Baumgardt \etal (2000) and Dias \etal (2002) based their computations on proper motions from the Tycho-2 catalogue
(H{\o}g \etal 2000). Kharchenko \etal (2004) used proper motions from the Tycho-2 catalogue,
the available photometry and the stars' positions in the cluster. Frinchaboy \etal (2008) used radial velocities
measured with the Hydra multi-object spectrographs, proper motions from the Tycho-2 catalogue, and the 
angular distance of the stars from cluster center. All these papers concern stars brighter than $V=13$ mag and 
altogether list 48 stars for which the membership probability, $P$, is higher than 60\,\%. 

We note, however, that Frinchaboy \etal (2008) perform separate computations of cluster membership from
proper motions, $P_{\rm pm}$, and from radial velocities, $P_{\rm rv}$. Since for many stars these values
differ significantly from each other, and because the final value of $P$ computed by Frinchaboy \etal (2008) 
is a product of $P_{\rm pm}$ and $P_{\rm rv}$, these authors find only one star for which $P >90$\%, 
two, for which $P\simeq 50$\%, one, for which $P\simeq 20$\%, ten, for which $P\le 10$\%, and 38, for which $P=0$. 
These differences between $P_{\rm pm}$ and $P_{\rm rv}$ computed by Frinchaboy \etal (2008), differences 
between $P_{\rm pm}$ derived by Frinchaboy \etal (2008) and by Dias \etal (2002) from the same data (where
$P_{\rm pm}$ given by Frinchaboy \etal (2008) is always lower than $P_{\rm pm}$ computed 
Dias \etal (2002)), and the very low number of stars classified by Frinchaboy \etal (2008) as cluster members 
is very unexpected. Because the origin of these discrepancies is not clear and their study is beyond the scope of this paper,
in the following Sections we will take with care the probabilities computed by Frinchaboy \etal (2008) and refer only to the 
results obtained by Dias \etal (2002) and Kharchenko \etal (2004).

So far, the cluster has not been a subject of a variability search. The Hipparcos Catalogue (ESA 1997) 
classifies two stars, HIP\,98610 and 98793, as unsolved variables $(U)$, and the Tycho Catalogue (ESA 1997),
five, HD\,190044, 190465, 190657, 190966, and TYC2-3162-00893-1, as suspected variables $(W)$.

\begin{figure}[htb]
\includegraphics{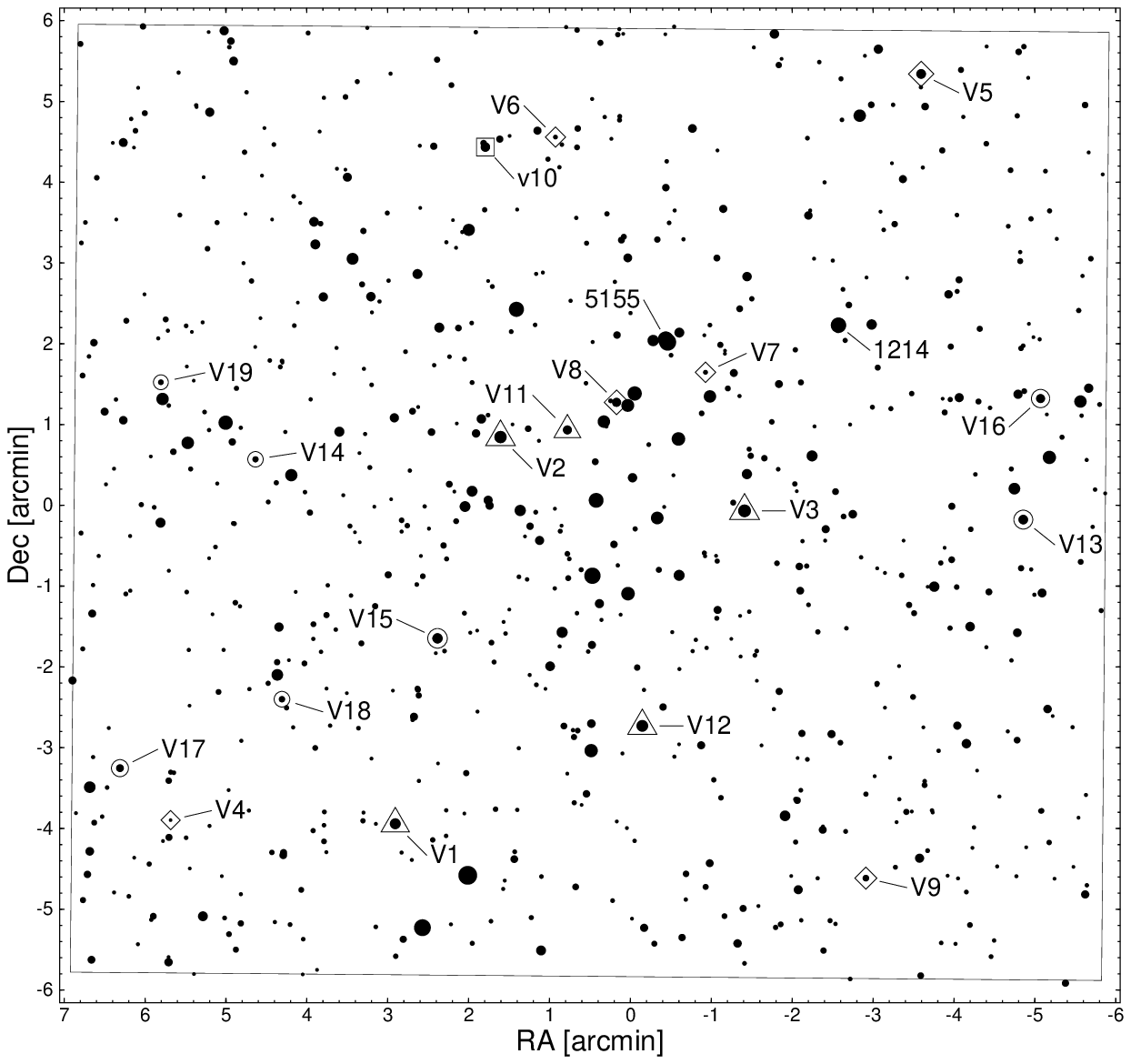}
\FigCap{The finding chart for the observed field in NGC\,6866. Only stars 
brighter than 18 mag in $V$ are shown. $\delta$ Sct and $\gamma$ Dor stars 
are indicated with triangles, ellipsoidal and EW stars, with diamonds, 
the EA system, with a square, and irregular variables, with circles.
The two blue--stragglers, NGC\,6866-1214 and -5155, are indicated with their WEBDA numbers.
The equatorial coordinates of point $(0,0)$ are equal to $\alpha _{\rm 2000} = 
20^{\rm h}03^{\rm m}55^{\rm s}$, $\delta _{\rm 2000} = 44^{\circ}09^{\prime}30^{\prime\prime}$.
}
\end{figure}

The paper is organized as follows. In Sect.\ 2, we give an account of the observations and reductions. 
In Sect.\ 3, we compute the cluster membership of stars in the field of NGC\,6866 and discuss the new variables.
In Sect.\ 4, we construct the color-magnitude diagram for NGC\,6866 and discover two 
blue-stragglers, the first such objects in the cluster. In Sect.\ 5 we discuss the issue of the age 
and metallicity of the open clusters in which $\gamma$ Dor stars can exist. Sect.\ 6 contains a summary.

The numbering system of stars in NGC\,6866 used in this paper is adopted from the 
WEBDA\footnote{http://www.univie.ac.at/webda/webda.html} database.

\section{Observations and Reductions}

\begin{figure}[htb]
\includegraphics{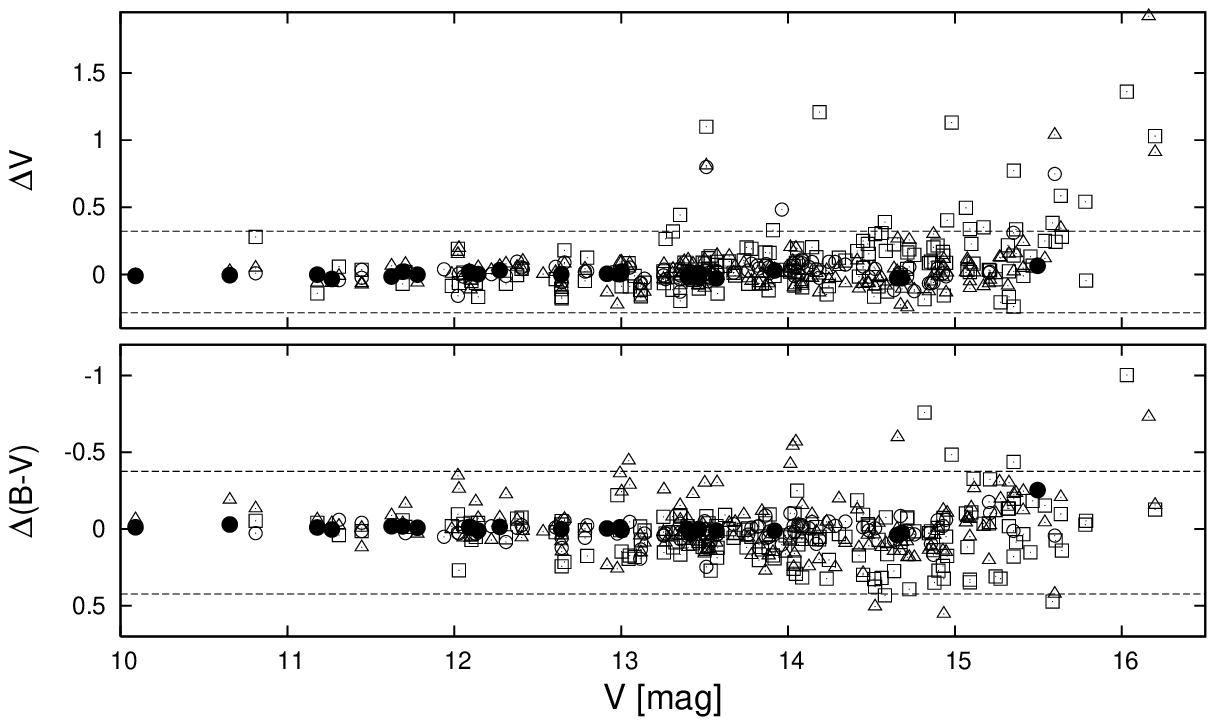}
\FigCap{The differences between the $V$ magnitudes and $(B-V)$ colors from this paper and 
those from the photoelectric and photographic photometry of Hoag \etal (1961) (filled and 
open circles, respectively), the photographic photometry of Barkhatova \& Zakharova (1970)
(triangles), and the photographic photometry of Hidajat \& Sutantyo (1972) (squares).
The dashed lines indicate the limits adopted for the $3\sigma$--clipping algorithm. 
}
\end{figure}

Our observations were carried out at the Bia\l{}k\'ow Observatory of the Wroc\l{}aw 
University on 14 nights between April 27 and July 21, 2007.
We used a 60-cm Cassegrain telescope equipped with Andor DW432-BV
back-illuminated CCD camera to observe one 12.8${}\times{}$11.7 arcmin$^{2}$
field in NGC\,6866. We used $BV$ and $I_{\rm C}$ filters of the 
Johnson-Kron-Cousins $UBV(RI)_{\rm C}$ photometric system and collected 
413 CCD frames in the $B$ filter, 470, in $V$, and 469, in $I_{\rm C}$. The 
exposure times were equal to 60 -- 100\,s, depending on the filter and the weather 
conditions.

In the pre-processing of the frames, the bias and dark frames were subtracted, 
the flat-field correction was applied, and the frames in the $I_{\rm C}$ 
filter were corrected for the fringing pattern. For all stars in the field the 
instrumental magnitudes were computed using the DAOPHOT profile-fitting software 
(Stetson 1987). The reductions were done as described by Jerzykiewicz \etal (1996).

In the $I_{\rm C}$-band reference frame of the field, we identified 3798 stars
for which we obtained meaningful photometry. The finding chart for the field we
observed is shown in Fig.\ 1. To avoid crowding, we show only stars brighter than 
18 mag in {\it V.}

\MakeTable{lrcrrrcrr}{12.5cm}{Mean differences between the $V$ magnitudes and $(B-V)$ colors 
from this paper and from the literature. 
}
{\hline\noalign{\smallskip}
& \multicolumn{4}{l}{$\Delta V$ [mag]}&\multicolumn{4}{l}{\hspace*{7pt}$\Delta (B-V)$ [mag]}\\
                              & mean & $\sigma$& $N$ & $N_{\rm rej}$ & mean & $\sigma$ & $N$ & $N_{\rm rej}$\\
\noalign{\smallskip}\hline\noalign{\smallskip}
photoelectric: \\
Hoag \etal (1961)            &0.002 &0.023 & 24 & 0 &  $-$0.011& 0.054 &  24& 0\\
\noalign{\smallskip}\hline\noalign{\smallskip}
photographic:\\ 
Hoag \etal (1961)            &0.018 &0.065 &  81& 3 &  0.015 & 0.065 &  84& 0\\
Barkhatova and Zakharova (1970)&0.006 &0.094 & 150& 5 & $-$0.001 & 0.148 & 147& 8\\
Hidajat and Sutantyo (1972)    &0.037 &0.128 & 134&17 &  0.059 & 0.146 & 145& 6\\
\noalign{\smallskip}\hline
}

We derived the differential photometry on a frame-to-frame basis, so that
the instrumental photometry for each frame had to be shifted to a common 
magnitude scale defined by a selected reference frame as described by Kopacki
\etal (2008).
Then, the average $v$ instrumental magnitudes and the $(b-v)$ colors-indices of 
all stars were transformed to the standard system using following equations 

\begin{eqnarray}
   V-v & = &-0.107\,(b-v)-1.232,\mbox{\hspace*{1cm}} \sigma={}0.019\mbox{\, mag,} \nonumber \\
   B-V & = &+1.276\,(b-v)-0.160,\mbox{\hspace*{1cm}} \sigma={}0.016\mbox{\, mag,} \nonumber
\end{eqnarray}
\noindent
that were obtained by the method of least squares from $24$ bright stars in common with 
Hoag \etal (1961). Here, $\sigma$ denotes standard deviation of the fit, $V$, the standard magnitude,
and $(B-V)$, the standard color index.

In Fig.\ 2, we plot the differences between the $V$ magnitudes and $(B-V)$ colors from 
this paper and those from Hoag \etal (1961), Barkhatova \& Zakharova (1970), and Hidajat \& Sutantyo (1972). 
In Table 1, we give the mean values of these differences which we calculated using a $3\sigma$-clipping 
algorithm for rejecting outlying data points, the standard deviations of the sample, 
$\sigma$, the number of points used in calculations, $N$, and the number of the rejected data 
points, $N_{\rm rej}$. The agreement between our measurements and those given in the literature is reasonably good.
The $V$ magnitudes and $(B-V)$ colors from this paper are given in Table 3, available
in electronic form from the Acta Astronomica Archive.

\def\mp{\omit\hfil--\hfil}
\vbox{\noindent\hbox to\hsize{\hss T a b l e 2\hss}\par
\vskip0.3\baselineskip\small
\noindent\hfil Photometric data for variable stars in \hbox{NGC\,6866}\par
\vskip\baselineskip
\normalbaselineskip=\baselineskip
\setbox\strutbox=\hbox{\vrule height0.7\normalbaselineskip%
 depth0.3\normalbaselineskip width0pt}
 \hbox to\hsize{\hfill\vbox{\tabskip=1pt \TableFont
  \halign{%
   \hfil#\hfil\tabskip=5.5pt&
   \hfil#\tabskip=7.5pt&
   #\hfil\tabskip=5.5pt&
   \hfil#\hfil&
   \hfil#\hfil&
   \hfil#\hfil&
   \hfil#\hfil&
   \hfil#\hfil&
   \hfil#\hfil&
   \hfil#\hfil&
   #\hfil%
   \tabskip=1pt\cr
   \noalign{\hrule\vskip3pt}
 Var& No & Type& $\alpha_{2000}$& $\delta_{2000}$& 
 $V$& $B$$-$$V$& $\Delta B$& $\Delta V$& $\Delta{i_{\rm C}}$& P\cr
   \noalign{\vskip1pt}
      &       &       & [$^{\rm h}$ $^{\rm m}$ $^{\rm s}$]&
                [$^\circ$ $^\prime$ $^{\prime\prime}$]&
 [mag]& [mag]& [mag]& [mag]& [mag]& [d]\cr
   \noalign{\vskip3pt\hrule\vskip3pt}
  V1& 3213& $\delta$ Sct& 20 04 11.19& 44 05 33.5& 12.976 & 0.319 &  0.009 &
0.008 & 0.008 & 0.066677\cr
  V2& 2105& $\delta$ Sct& 20 04 03.96& 44 10 20.7& 12.376 & 0.302 &  0.006 &
0.006 & 0.007 & 0.072465\cr
  V3&   29& $\delta$ Sct& 20 03 47.13& 44 09 25.9& 12.222 & 0.302 &  0.018 &
0.015 & 0.012 & 0.106414\cr
    &     &             &            &           &        &       &  0.009 &
0.011 & 0.008 & 0.120744\cr
    &     &             &            &           &        &       &  0.004 &
0.006 & 0.003 & 0.085202\cr
\noalign{\vskip3pt}
  V4&   --& W UMa&        20 04 26.66& 44 05 36.2&  --    & --    &  --    &
--    & 0.601 & 0.262524\cr
  V5& 1311& W UMa&        20 03 34.93& 44 14 50.4& 13.524 & 0.443 &  0.030 &
0.047 & 0.055 & 0.321742\cr
  V6&   --& W UMa&        20 04 00.17& 44 14 03.5& 17.331 & 0.886 &  0.477 &
0.467 & 0.390 & 0.366528\cr
  V7&   --& W UMa&        20 03 49.82& 44 11 08.8& 17.196 & 0.862 &  0.384 &
0.341 & 0.308 & 0.41501\cr
  \noalign{\vskip3pt}
  V8&   39& Ell&          20 03 55.96& 44 10 46.5& 13.868 & 0.430 &  0.047 &
0.029 & 0.024 & 0.6222\cr
  V9&  133& Ell&          20 03 38.79& 44 04 53.1& 15.543 & 0.773 &  0.213 &
0.193 & 0.169 & 0.43414\cr
  \noalign{\vskip3pt}
 V10&   --& EA         &  20 04 05.01& 44 13 56.1& 13.628 & 0.606 &  0.123 &
0.092 & 0.093 & 1.916\cr
  \noalign{\vskip3pt}
 V11&   21& $\gamma$ Dor& 20 03 59.34& 44 10 26.0& 13.919 & 0.431 &  0.100 &
0.069 & 0.043 & 0.8057\cr
    &     &             &            &           &        &       &  0.036 &
0.036 & 0.018 & 0.9060\cr
 V12& 4108& $\gamma$ Dor& 20 03 54.18& 44 06 46.2& 12.644 & 0.228 &  0.020 &
0.016 & 0.013 & 0.7077\cr
  \noalign{\vskip3pt}
 V13&   20& Irr&          20 03 27.92& 44 09 19.4& 13.571 & 0.431 &  0.146 &
0.121 & 0.079 & --\cr
 V14&   --& Irr&          20 04 20.84& 44 10 04.2& 15.796 & 1.180 &  0.279 &
0.176 & 0.113 & --\cr
 V15& 3110& Irr&          20 04 08.29& 44 07 51.2& 13.080 & 2.043 &  0.133 &
0.097 & 0.047 & --\cr
 V16& 1220& Irr&          20 03 26.72& 44 10 49.3& 13.751 & 2.120 &  0.223 &
0.083 & 0.035 & --\cr
 V17&  150& Irr&          20 04 30.15& 44 06 14.8& 14.514 & 2.049 &  0.284 &
0.218 & 0.076 & --\cr
 V18&   --& Irr&          20 04 19.00& 44 07 06.0& 15.570 & 2.108 &  0.770 &
0.395 & 0.148 & --\cr
 V19&   --& Irr&          20 04 27.38& 44 11 01.4& 16.231 & 2.209 &  1.048 &
0.569 & 0.289 & --\cr
  \noalign{\vskip3pt\hrule}
  }}\hfill}}

\section{Variable Stars}

For each star, the Fourier spectrum and an AoV periodogram (Schwar\-zen\-berg-Czer\-ny 1989) 
were computed in the frequency range from 0 to 50 d$^{-1}$. These two methods were used because the 
former is useful for stars of which the brightness varies sinusoidally, while the latter, for eclipsing 
binaries. Then, we calculated the signal-to-noise ratio, $S/N$, of the highest peak in each spectrum and 
checked by eye the phase-diagrams corresponding to stars with $S/N\ge 4$ for the presence of a 
periodic variability or eclipses.
In Fig.\ 3, we plot the $S/N$ ratio of the highest peak in the frequency spectrum (in the figure we show only 
the 0--20 d$^{-1}$ part of the spectrum) against its frequency for all stars in the field. 

\begin{figure}[htb]
\includegraphics{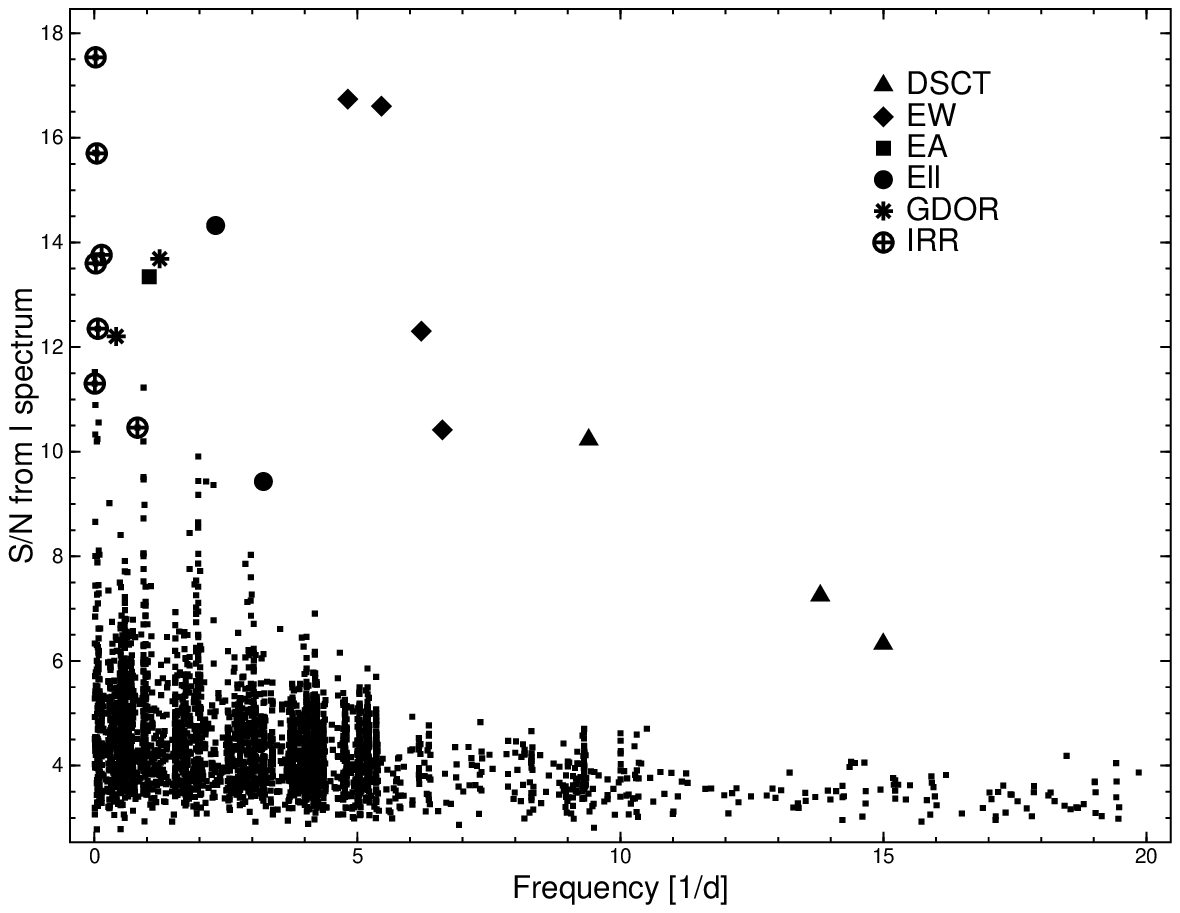}
\FigCap{The signal-to-noise ratio (S/N) of the highest peak in the frequency spectrum of the 
I--filter data for each star in the field we observed. Variable stars are indicated with the following 
symbols: $\delta$ Sct, triangles, $\gamma$ Dor, asterisks, W\,UMa, diamonds, the EA system,
a square, ellipsoidal variables, bullets, and irregular variables, encircled plus signs.
}
\end{figure}

We discovered 19 variable stars which we classified on the basis of the length 
of the detected period and the shape of the light-curve. We designated these stars V1 
through V19. The stars are listed in Table 2, where we give their designation, 
the number from WEBDA, type of variability, equatorial coordinates, the mean brightness in 
$V$ and the mean color-index $(B-V)$ for all stars but the faintest V4 for which we 
do not compute the standard magnitudes, the range of variability in $B$, $V$ and the instrumental 
$i_{\rm C}$ filter, and for the periodic variables, the period(s), $P$. 

The variability of HIP\,98793, NGC\,6866-5, classified in the Hipparcos Catalogue as an unsolved variable, 
has not been confirmed in our data. We also do not confirm variability of this star in Hipparcos
$Hp$ magnitudes. The remaining suspected and unsolved variables listed in the 
Hipparcos and the Tycho Catalogues were not included in our field of view.

\begin{figure}[htb]
\includegraphics{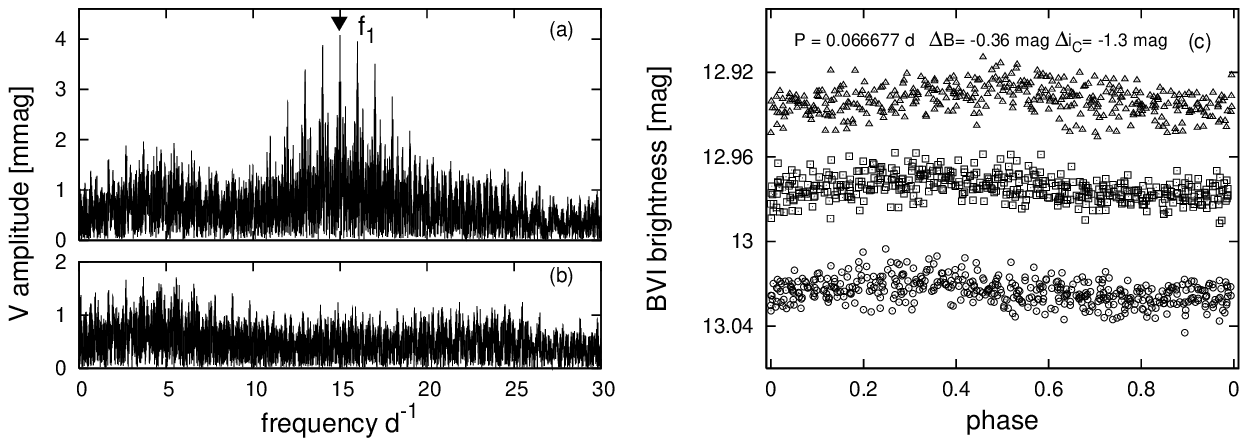}
\FigCap{Fourier spectra of the $\delta$ Sct star V1:
 $(a)$ for original $V$-filter observations,
 $(b)$ after prewhitening with frequency $f_1=14.9976$ d$^{-1}$ (the ordinate scale is the same in both panels);
 $(c)$ the light curves of V1 in $B$ (triangles), $V$ (squares) and the instrumental $i_{\rm C}$ filter (circles).
 $\Delta B$ and $\Delta i_C$ are magnitude shifts applied to $B$ and $i_C$ data, respectively.
}
\end{figure}

\begin{figure}[htb]
\includegraphics{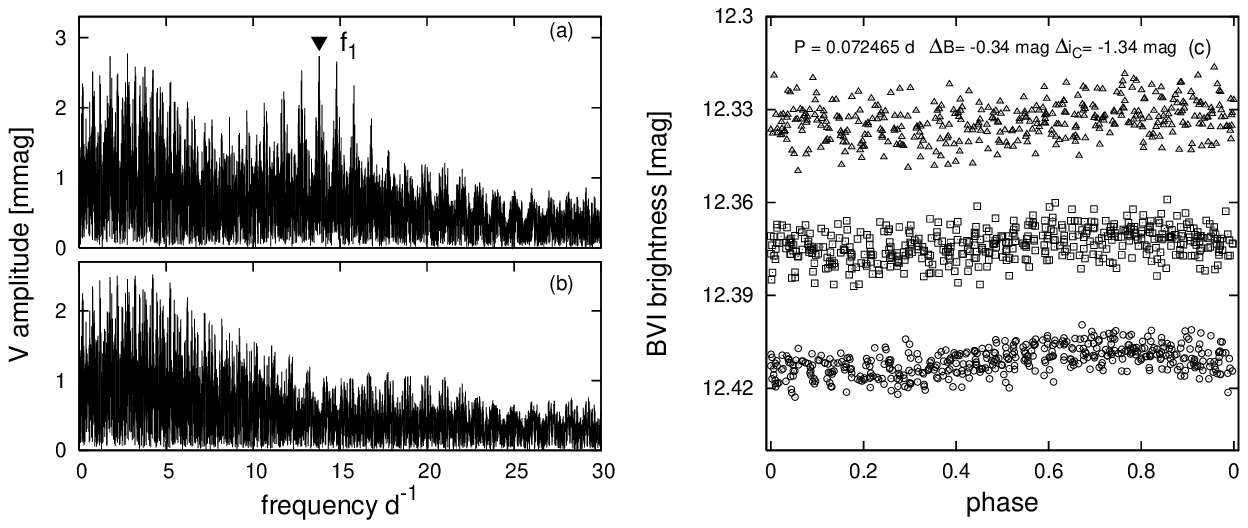}
\FigCap{The same as in Fig.\ 4 but for the $\delta$ Sct star V2. $(a)$ The highest peak occurs at the frequency 
$f_1= 13.7999$ d$^{-1}$  (the ordinate scale is the same in both panels).
$(c)$ The light curves of V2 in $B$ (triangles), $V$ (squares) and the instrumental 
$i_{\rm C}$ filter (circles).
$\Delta B$ and $\Delta i_C$ are magnitude shifts applied to $B$ and $i_C$ data, respectively.
}
\end{figure}

\subsection{Cluster Membership}

Only two of the 19 variables discovered in this paper have the cluster membership probability 
computed by Dias \etal (2002) or Kharchenko \etal (2004). Therefore, we used the proper motions
from R\"oser \etal (2008) to compute the probability of membership in the cluster 
for stars from our field. First, we used stars that fall into the cluster area centered at 
$\alpha _{2000} =20^{\rm h}.065$, $\delta _{2000}=44^{\circ}.16$ and limited by the radius of 
$0^{\circ}.14$ adopted from Kharchenko \etal (2005) to compute the mean proper motion of the cluster. The 
computations were done iteratively; we used a $\sigma$--clipping algorithm and rejected stars
having the proper motion that differed from the mean proper motion by more than $3\sigma$. 
The resulting value, 
$\mu _{\alpha}\cos\delta = -3.86\pm0.16$, $\mu_{\delta} = -4.63\pm0.17$ mas/yr (the uncertainty 
given here is the r.m.s.\ error), agrees well 
with the mean cluster proper motion computed by Dias \etal (2002), $\mu _{\alpha}\cos \delta = 
-3.33\pm 2.93$, $\mu _{\delta} = -5.03\pm 2.93$ mas/yr, and Kharchenko \etal (2005), $\mu 
_{\alpha}\cos\delta = -3.48\pm 0.40$, $\mu _{\delta} = -5.80\pm 0.38$ mas/yr. The agreement 
with the values given by Frinchaboy \etal (2008), $\mu _{\alpha}\cos\delta = -5.5\pm1.2$, $\mu_{\delta} =
-8.0\pm1.1$, is less satisfactory. We note that the uncertainties 
given by Dias \etal (2002) and Frinchaboy \etal (2008) are standard deviations while 
Kharchenko \etal (2005) give the r.m.s.\ errors.

\begin{figure}[htb]
\includegraphics{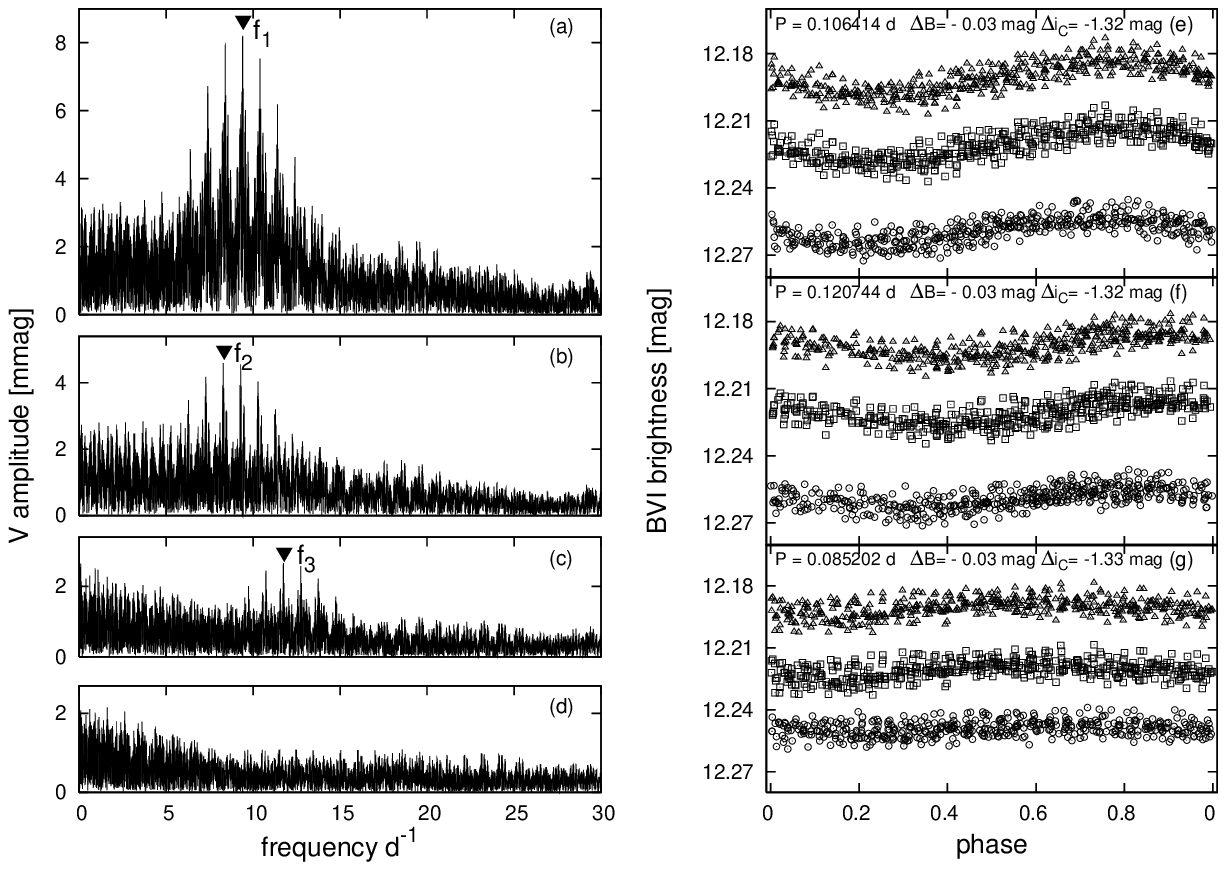}
\FigCap{Similar as in Fig.\ 4 but for the $\delta$ Sct star V3. The highest peaks occur at 
$(a)$ $f_1= 9.3973$ d$^{-1}$, $(b)$ $f_2= 8.2820$ d$^{-1}$ and $(c)$ $f_3= 11.7368$ d$^{-1}$
(the ordinate scale is the same in all panels.)
$(e)$ The light curves of V3 in $B$ (triangles), $V$ (squares) and the instrumental $i_{\rm C}$ filter (circles),
prewhitened with the frequencies $f_2$ and $f_3$, and phased with
the frequency $f_1$; $(f)$ -- $(g)$ the same as in $(e)$ but for the frequencies $f_2$ and $f_3$.
$\Delta B$ and $\Delta i_C$ are magnitude shifts applied to $B$ and $i_C$ data, respectively.
}
\end{figure}

Then, following Kharchenko \etal (2005), for each star for which the proper motion has been measured, we 
computed the cluster membership probability, $P,$ as a measure of its deviation from the mean 
proper motion of the cluster, $d$. In the result, 23 stars with $d \le \sigma$ (where $\sigma$ is the standard 
deviation of the proper motions of the sample), i.e., $P > 60\,\%$, we classified as 
most probable cluster members, 20 stars 
with $1\sigma \le d < 2 \sigma$, i.e., $14\% \le P \le 60\%$, as possible members, 11 
stars with $2\sigma \le d < 3 \sigma$, i.e., $1\% \le P \le 14\%$, as possible field stars, and 
19 stars with $d<3 \sigma$, i.e., $P<1\%$, as definite field stars. The cluster membership 
probability computed in this paper is listed in the last column of Table 3.

Our computations show that the $\delta$ Sct stars V1, V2 and V3, and the 
$\gamma$ Dor star V12 belong to the cluster. The eclipsing binary, V10, and one of 
the irregular variables, V13, are definite field stars. For the other irregular variable 
V15, the cluster membership probability is 30\,\% so that it could be a cluster member. However, 
its location on the color-magnitude diagram far to the red of the main sequence (see Fig.\ 11) 
rules out that possibility.

\begin{figure}[htb]
\includegraphics{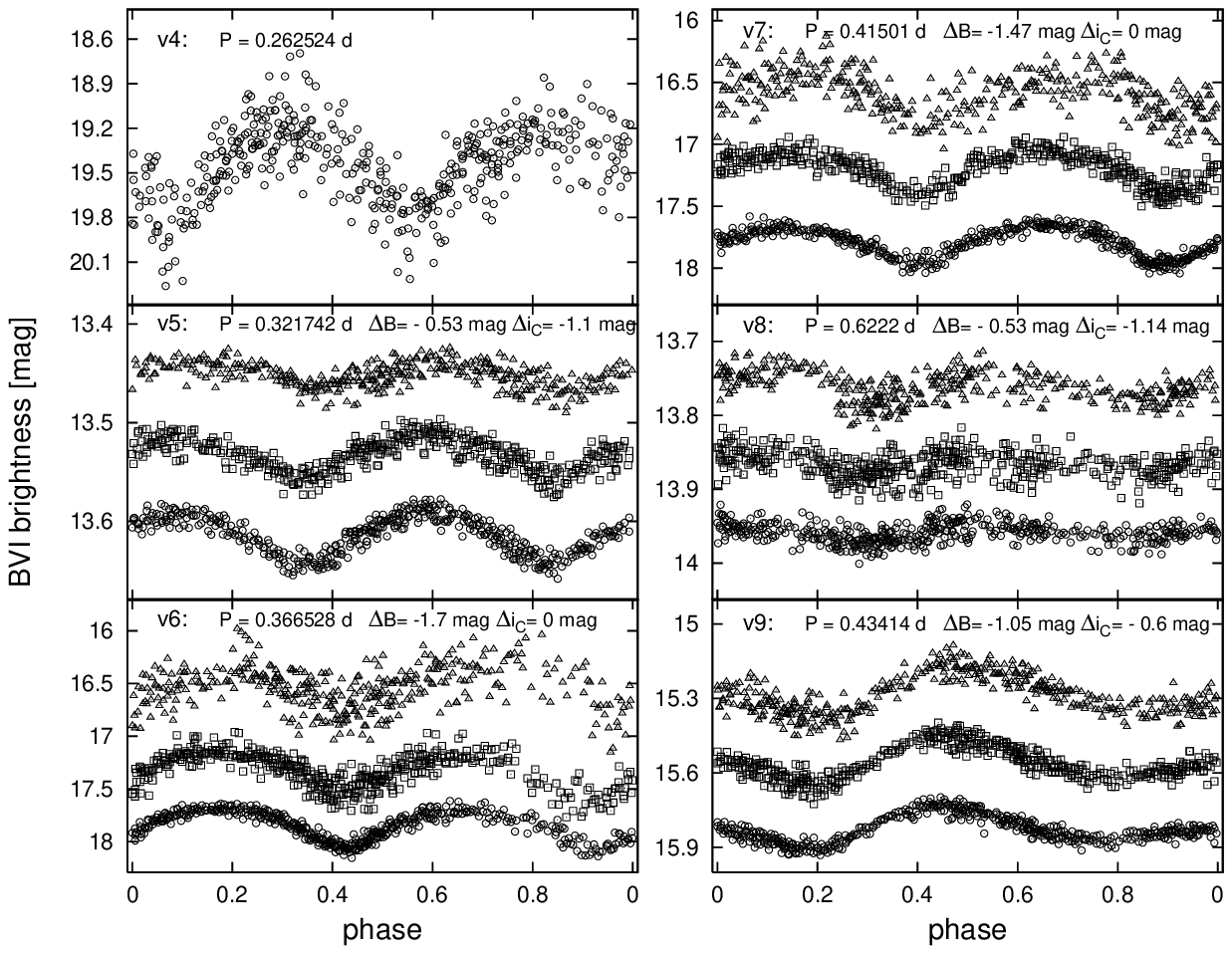}
\FigCap{The $B$ (triangles), $V$ (squares) and the instrumental $i_{\rm C}$ filter (circles)
 light curves of the W Ursae Majoris systems, V4, V5, V6 and V7, and the ellipsoidal stars, V8 and V9.
 $\Delta B$ and $\Delta i_C$ are magnitude shifts applied to $B$ and $i_C$ data, respectively.
}
\end{figure}

\subsection{$\delta$ Sct stars}
We discovered two monoperiodic $\delta$ Sct stars, V1 and V2, 
and one multiperiodic $\delta$ Sct star, V3. V1 and V2 vary with the frequencies  
$f_1=14.9976$ and $f_1= 13.7999$ d$^{-1}$, respectively. V3 shows three frequencies, $f_1= 9.3973$,
$f_2= 8.2820$ and $f_3= 11.7368$ d$^{-1}$. All these stars lie on the main sequence in the cluster's 
color-magnitude diagram (see Fig.\ 11) and for all the probability of cluster membership is high: 
80\,\%, 71\,\% and 87\,\% for V3 computed by, respectively, Dias \etal (2002), Kharchenko \etal (2004) 
and in this paper, and 72\,\% and 94\,\% computed for V1 and V2 in this paper. Therefore, 
we consider all these stars to be very probable members of NGC\,6866.

We show the amplitude spectra of V1 and V2 in Figs.\ 4 and 5, respectively, and that of V3, in Fig.\ 6.
The amplitude spectra of the residuals, plotted in Figs.\ 4$b$, 5$b$ and 6$d$ show only 
noise. In Figs.\ 4$c$, 5$c$ and Fig.\ 6$e$-$g$, we plot the phase diagrams constructed 
from $B$, $V$ and the instrumental $i_{\rm C}$ magnitudes. As can be
seen from these figures, the stars show sinusoidal variations of the brightness in all filters. 

\subsection{W\,UMa and Ell stars}

We discovered four W\,UMa type variables, V4, V5, V6 and V7, and two ellipsoidal variables, 
V8 and V9. We show their light curves in  Fig.\ 7. For V4, we show only $i_{\rm C}$-- filter light curve
because in the other filters the scatter masks the variation.

\subsection{The eclipsing binary}

Our time-series of V10 shows two minima which have the same shape of the ingress (the egress
was observed only partially) and which we interpret as the same type of minimum. 
The $BVi_{\rm C}$ light-curves, phased with $P_{\rm orb} = 1.916$d, are plotted in Fig.\ 8; 
we have checked that neither $0.5 P_{\rm orb}$ nor any other period fits the data. 
The observed minimum is flat. The brightness between eclipses is not constant.

\begin{figure}[htb]
\includegraphics{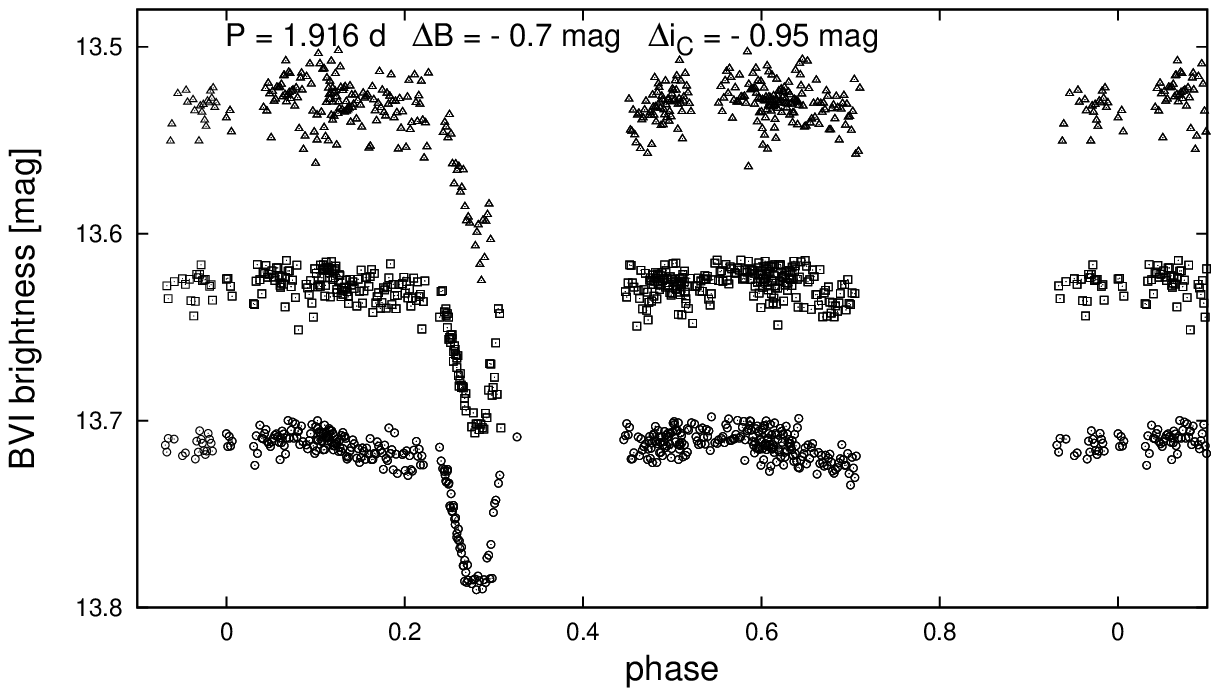}
\FigCap{The $B$ (triangles), $V$ (squares) and the instrumental $i_{\rm C}$ filter (circles)
 light curves of the eclipsing binary V10. $\Delta B$ and $\Delta i_C$ are magnitude shifts applied to 
 $B$ and $i_C$ data, respectively.
}
\end{figure}

The proper motion of V10, $\mu _{\alpha} \cos \delta = -37.75\pm 14.13$, $\mu 
_{\delta} = -54.14\pm 14.3$ mas/yr (R\"oser \etal 2008), is around ten times higher 
than the mean proper motion of the cluster. The cluster membership probability of V10 computed 
in this paper is lower than $1\,\%$. Having reached the conclusion that V10 is a definite field
star and keeping in mind its high proper motion, we suspected that V10 is a high-velocity
metal-deficient Population II star. 

Because our data do not allow deciding whether the primary or the secondary minimum has been 
observed, we calculated the range of magnitudes of the primary and the secondary component of V10
which give the observed magnitude at the quadrature, $V=13.33$, $B=13.76$ mag, and at the 
minimum, $V=13.41$, $B=13.86$ mag. Because the orbital inclination of the system has not been measured, 
in our calculations we adopted $i=90^{\circ}$. The calculated magnitudes cover all the possible combinations
of the magnitudes of the components, i.e., from the secondary component being so faint 
that it does not contribute to the mean brightness of the system outside eclipses to both components 
having the same brightness. The resulting $B$ and $V$ magnitudes of the primary, $1$,
and the secondary, $2$, components fall into the following ranges: $m_{V,1} \in (13.33;14.09)$, which corresponds
to $m_{V,2}\in (+\infty;14.09)$, and $m_{B,1}\in (13.76;14.52)$, which corresponds to 
$m_{B,2}\in (+\infty;14.52)$ mag.

\begin{figure}[htb]
\includegraphics{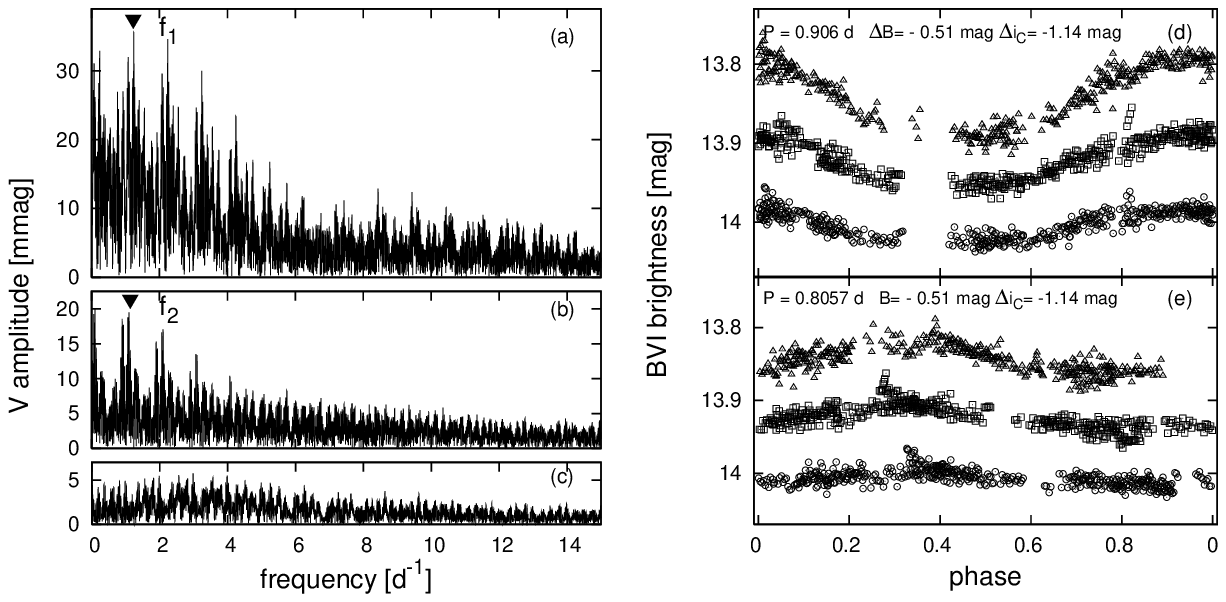}
\FigCap{Similar as in Fig.\ 4 but for the $\gamma$ Dor star V11. The highest maximum occurs at
 $(a)$ $f_1= 1.2412$ d$^{-1}$ and $(b)$ $f_2= 1.1038$ d$^{-1}$  (the ordinate scale is the same in all panels.)
$(d)$--$(e)$ The $B$ (triangles), $V$ (squares) 
 and the instrumental $i_{\rm C}$ filter (circles) light curves of V11. $\Delta B$ and $\Delta i_C$ 
 are magnitude shifts applied to $B$ and $i_C$ data, respectively.

}
\end{figure}

\begin{figure}[htb]
\includegraphics{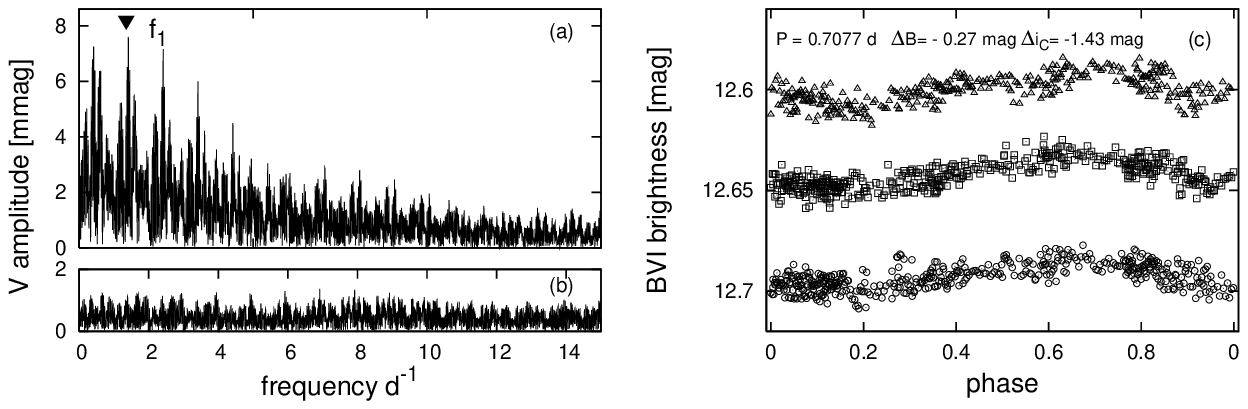}
\FigCap{The same as in Fig.\ 4 but for the $\gamma$ Dor candidate V12. $(a)$ The highest maximum occurs at the 
frequency $f_1 = 1.4130\,\rm d^{-1}$  (the ordinate scale is the same in both panels.)
$(c)$ The $B$ (triangles), $V$ (squares)
 and the instrumental $i_{\rm C}$ filter (circles) light curves of V11. $\Delta B$ and $\Delta i_C$
 are magnitude shifts applied to $B$ and $i_C$ data, respectively.

}
\end{figure}

Since we aimed at calculating the lower limit of the tangential velocity of V10, we assumed that the star is a 
not reddened metal-deficient dwarf, and that the brightness of its primary component is $V=13.33$; any change in
our assumptions, e.g., an increase of $E(B-V)$, higher magnitude of the primary component, or a higher 
metallicity (i.e., brighter absolute magnitude), would result in a higher tangential velocity of the star.
We calculated the star's absolute magnitude, $M_V= 4.9$, from $M_V$--$(B-V)$ relation of
Karata\c{s} \& Schuster (2006) for metal-poor stars. Then, we calculated the distance to the star, $r= 489$\,pc, 
and its tangential velocity, 
152 km/s. Since the calculated value is higher than 100 km/s, which is the lower limit of tangential velocity of 
high-velocity stars (see, e.g., Lee (1984)), we classify V10 as a new high-velocity star in the field of NGC\,6866.

\subsection{$\gamma$ Dor stars}

Two stars in NGC\,6866 show variablity on time-scale of a day; V11, which shows 
two frequencies, $f_1 = 1.2412$ d$^{-1}$ and $f_2= 1.1038$ d$^{-1}$, and V12, 
which shows one frequency, $f_1 = 1.4130$ d$^{-1}$. For both stars, the amplitudes are the 
highest in the $B$ filter and the smallest, in $i_{\rm C}$, and the light-curves are sinusoidal. 
The amplitude spectra and phase diagrams for V11 and V12 are shown in Figs.\ 9 and 10.

\begin{figure}[htb]
\includegraphics[totalheight=3in]{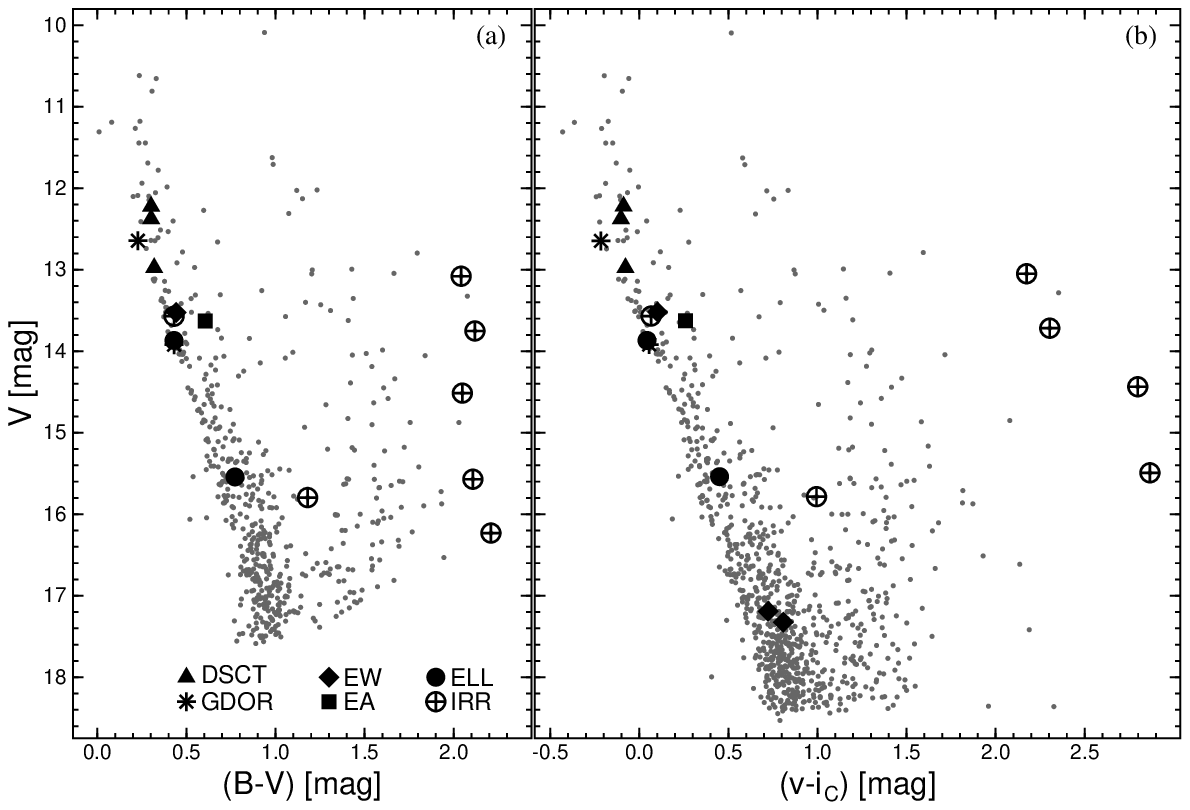} 
\FigCap{$(a)$ The standard $(B-V)$ vs.\ $V$ and $(b)$ the instrumental $(v-i_{\rm C})$ vs.\ $V$
color -- magnitude diagrams for NGC 6866. $\delta$ Sct stars are indicated with triangles, 
$\gamma$ Dor, with asterisks, W\,UMa, with diamonds, the eclipsing binary, 
with a square, ellipsoidal variables, with bullets, and irregular variables, with encircled plus signs.
Panel $b$ does not include V19 for which $(v-i_{\rm C}) = 4.48$ mag.
}
\end{figure}

\begin{figure}[htb]
\includegraphics[totalheight=3in]{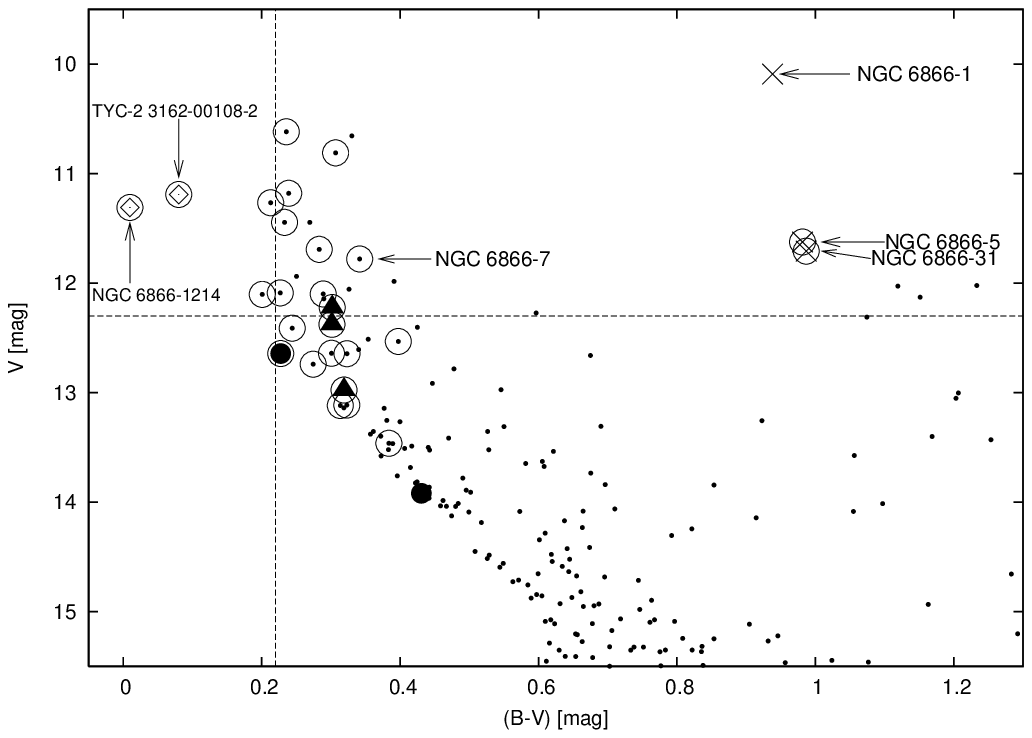}
\FigCap{A zoom of Fig.\ 11$a$. Circles indicate stars for which the membership probability 
is higher than 60\,\%. Blue--stragglers are indicated with open diamonds, red giants, with crosses.
$\delta$ Sct stars are indicated with triangles, $\gamma$ Dor stars, with bullets; the remaining 
variables are not included.
The intersection of the dashed lines indicates the turn-off point of the cluster, i.e., 
$m_v$ and $(B-V)$ of an A3V star in a distance of
1200 pc and reddened with $E(B-V)=0.14$ mag.  NGC\,6866--7 is a rejected blue--straggler candidate.
}
\end{figure}

The probability of membership of V12 in NGC\,6866 is 97\% or 66\,\% according to 
Kharchenko \etal (2004) and this paper, respectively. The cluster membership of V11 has not
been computed. However, as both stars fall on the cluster main sequence, we consider both 
V11 and V12 to be probable cluster members.

Having discovered the multiperiodic variability of V11, and taking into account the length of the detected 
periods, we classify the star as a $\gamma$ Dor -- type variable. The other star, V12, we classify as a $\gamma$ Dor 
suspect on the basis of the length of the detected period and the increase of the amplitude of its variability 
towards the shorter wavelengths.

In Sect.\ 5, we discuss the phenomenon of the occurrence of $\gamma$ Dor stars in open clusters in more detail.

\subsection{Irregular variables}

The irregular variables V13, V14, V15, V16, V17, V18 and V19 form the largest group of variable stars 
discovered in the field of NGC\,6866. We computed the probability of membership in the cluster for V13
and V15. V13, which falls on the cluster main sequence (see Fig.\ 11),
turned out to be a definite field star, and V15, which might be a probable cluster member, 
is excluded as such because of its position far to the red of the cluster main sequence.
Therefore, we consider all the irregular variables to be non-members.

\setcounter{table}{3}
\MakeTable{lrrl@{\hspace{30pt}}lrrl}{12.5cm}{The list of Galactic open clusters which host candidates for $\gamma$ Dor stars.
}
{\hline\noalign{\smallskip}
cluster & $N_{\gamma\, \rm Dor}$ & log age & $\rm [Fe/H]$ & cluster & $N_{\gamma\, \rm Dor}$ & log age & $\rm [Fe/H]$\\
\noalign{\smallskip}\hline\noalign{\smallskip}
NGC 581 & 2 {\hspace*{1pt}} {\tiny [1]} &    7.1 {\hspace*{1pt}} {\tiny [2]}&$-$0.85 {\hspace*{1pt}} {\tiny [3]} & 
NGC 6633& 2 {\tiny [61]}                &    8.6 {\tiny [25]}               &$-$0.30 {\tiny [22]} \\ \vspace{4pt}
        & \tiny (1 nm)                  &    7.3 {\hspace*{1pt}} {\tiny [2]}& {\hspace*{8pt}} --- &
        &                               &    8.8 {\tiny [38]}               &$+$0.06 {\tiny [39]} \\

NGC 659 & 3 {\hspace*{1pt}} {\tiny [4]} &    7.3 {\hspace*{1pt}} {\tiny [5]}& {\hspace*{8pt}} --- & 
NGC 6705& 2 {\tiny [40,41]}             &    8.3 {\tiny [12]}               &$+$0.07 {\hspace*{1pt}} {\tiny [9]}\\ \vspace{4pt}
        &                               &    7.6 {\hspace*{1pt}} {\tiny [6]}& {\hspace*{8pt}} --- &
        & \tiny (1 nm)                  &    8.4 {\tiny [42]}               &$+$0.21 {\tiny [43]}\\ 

NGC 1039& 2 {\hspace*{1pt}} {\tiny [7]} &    8.3 {\hspace*{1pt}} {\tiny [8]}&$-$0.29 {\hspace*{1pt}} {\tiny [9]}& 
NGC 6755& 4 {\tiny [44]}                &    7.7 {\tiny [12]}               &$-$0.03 {\hspace*{1pt}} {\tiny [3]} \\ \vspace{4pt}
        &                               &    8.4 {\hspace*{1pt}} {\tiny [8]} &$+$0.07 {\tiny [10]} &
        & \tiny (2 nm)                  &    8.2 {\hspace*{1pt}} {\tiny [3]} &$+$0.14 {\tiny [15]} \\

NGC 1245& 4 {\tiny [11]}                &    8.7 {\tiny [12]}&$-$0.05 {\tiny [14]}& 
NGC 6866& 2 {\tiny [45]}                &    8.4 {\tiny [46]}&$+$0.10 {\tiny [47]}\\ \vspace{4pt}
        &                               &    9.0 {\tiny [13]}&$+$0.10 {\tiny [15]}&
        &                               &    8.8 {\tiny [47]}& {\hspace*{8pt}} ---\\

NGC 1817& 1 {\tiny [16]}                &    8.9 {\tiny [17]}&$-$0.42 {\tiny [19]}& 
NGC 7086& 1 {\tiny [48]}                &    8.0 {\tiny [48]}& {\hspace*{8pt}} --- \\ \vspace{4pt}
        & \tiny (1 nm?)                 &    9.1 {\tiny [18]}&$+$0.18 {\tiny [15]} &
        &                               &    8.1 {\tiny [12]}& {\hspace*{8pt}} --- \\
 
NGC 2099& 2 {\tiny [20]}                &    8.2 {\tiny [21]}&$-$0.07 {\tiny [22]}& 
NGC 7762& 2 {\tiny [49,50]}             &    8.4 {\tiny [12]}& {\hspace*{8pt}} --- \\ \vspace{4pt}
        & \tiny (1 nm?)                 &    8.3 {\tiny [21]}&$+$0.18 {\tiny [23]}&
        &                               &    9.4 {\tiny [49]}& {\hspace*{8pt}} ---\\

NGC 2301& 2 {\tiny [24]}                &    8.2 {\tiny [25]}&$+$0.01 {\tiny [23]}& 
Pleiades& 2 {\tiny [37]}                &    7.9 {\tiny [31]}&$-$0.03 {\tiny [34]}\\ \vspace{4pt}
        &                               &    8.5 {\tiny [24]}&$+$0.06 {\tiny [25]}&
        &                               &    8.1 {\tiny [42]}&$+$0.11 {\tiny [12]}\\

NGC 2506&15 {\tiny [26]}                &    9.1 {\tiny [12]}&$-$0.57 {\tiny [28]}& 
$\alpha$ Per & 2 {\tiny [52]}           &    7.7 {\tiny [53]}&$-$0.05 {\tiny [51]}\\ \vspace{4pt}
        &                               &    9.3 {\tiny [27]}&$-$0.20 {\tiny [29]}&
        &                               &    8.0 {\tiny [54]}&$+$0.07 {\hspace*{1pt}} {\tiny [9]}\\

NGC 2516& 8 {\tiny [30]}                &    8.1 {\tiny [31]}&$-$0.42 {\tiny [15]}& 
Coma Ber& 1 {\tiny [55]}                &    8.5 {\tiny [56]}&$-$0.05 {\tiny [56]}\\ \vspace{4pt}
        &                               &    8.2 {\tiny [32]}&$+$0.02 {\tiny [15]}&
        &                               &    8.7 {\tiny [38]}&$-$0.03 {\tiny [58]}\\

NGC 2539& 1 {\tiny [31]}                &    8.8 {\tiny [33]}&$-$0.20 {\tiny [33]}& 
Praesepe& 1 {\tiny [52]}                &    8.8 {\tiny [59]}&$+$0.08 {\tiny [22]}\\ \vspace{4pt}
        & \tiny (1 nm)                  &    8.9 {\tiny [33]}&$+$0.16 {\tiny [16]}&
        &                               &    8.9 {\tiny [52]}&$+$0.27 {\tiny [60]}\\ 

NGC 6231& 3 {\tiny [34]}                &    6.5 {\tiny [35]}&$-$0.13 {\tiny [36]}&\\
        &                               &    6.7 {\tiny [35]}&$+$0.26 {\hspace*{1pt}} {\tiny [3]}&\\
\noalign{\smallskip\smallskip}\hline\noalign{\smallskip}
\multicolumn{8}{l}{
{\begin{minipage}{12cm}
\tiny [1] Wyrzykowski \etal 2002, [2] Sanner \etal 1999, [3] Tadross 2003, [4] Pietrzy\'nski \etal 2001, [5] Phelps and Janes 1994,
[6] Lata \etal 2002, [7] Krisciunas and Crowe 1996, [8] Jones and Prosser 1996, [9] Cameron 1985, [10] Schuler \etal 2003,
[11] Pepper and Burke 2006, [12] WEBDA, [13] Subramaniam 2003, [14] Burke \etal 2003, [15] Gratton 2000, [16] Arentoft \etal 2005,
[17] Harris and Harris 1977, [18] Balaguer-Nu\`n\~ez \etal 2004, [19] Friel and Janes 1993, [20] Hartman \etal 2008, [21] Kalirai
\etal 2001, [22] Janes 1979, [23] Piatti \etal 1995, [24] Kim \etal 2001b, [25] Chen \etal 2003, [26] Arentoft \etal 2007, [27]
Kim \etal 2001a, [28] Friel and Janes 1993, [29] Carretta \etal 2004, [30] Zerbi \etal 1998, [31] Choo \etal 2003, [32] Bonatto
and Bica 2005, [33] Clari\'a and Lapasset 1986, [34] Arentoft \etal 2001, [35] Baume \etal 1999, [36] Kilian \etal 1994, [37] 
Mart\'in and Rodr\'iguez 2002a, [38] Lyng\aa 1987, [39] Santos \etal 2009, [40] Koo \etal 2007, [41] Hargis \etal 2005, [42]
Magrini \etal 2009, [43] Thogersen \etal 1993, [44] Ciechanowska \etal 2007, [45] this paper, [46] Lindoff 1968, [47] Loktin
\etal 1994, [48] Rosvick and Robb 2006, [49] Maciejewski \etal 2008, [50] Szabo 1999, [51] Boesgaard and Friel 1990, [52] Mart\'in 
and Rodr\'iguez 2002b, [53] Makarov 2006, [54] Stauffer \etal 1999, [55] Mart\'in 2003, [56] Sandage 1958, [57] Friel and Boesgaard 1992,
[58] Cayrel \etal 1988, [59] Bouvier \etal 2001, [60] Pace \etal 2008, [61] Mart\'in and Rodr\'iguez 2001.
\end{minipage}
}
}
}

\section{The Color -- Magnitude Diagram}

In Fig.\ 11$a$ and $b$, we show $V$ vs.\ $(B-V)$ and $V$ vs.\ the instrumental 
$(v-i_{\rm C})$ color-magnitude diagrams for stars in the field of NGC\,6866 
for which we have reliable photometry. The intersection of the dashed lines in Fig.\ 12$b$ indicates
the turn-off point of the cluster, A3, estimated by Johnson \etal (1961).

As can be seen from these figures, most
stars fall on the main sequence; the red giant clump is not visible. 
Indeed, in NGC\,6866 there are only four red giants listed by Mermilliod \etal 2008:
NGC\,6866-1, 5, 26 and 31. (We note, however, that NGC\,6866-1 has been classified by 
Sowell~(1987) to G5V.) We observed three of these stars; NGC\,6866-26 was outside of our 
field of view. 

\subsection{Blue--stragglers}

NGC\,6866 is at the age at which, according to De Marchi \etal (2006), it should host one or two blue--stragglers
(see Table 1 in De Marchi \etal 2006). Unfortunately, the only blue--straggler suspect, NGC\,6866-7 
(Hoag \etal 1961), has been rejected by Ahumada \& Lapasset 
(2007). In the color-magnitude diagram in Fig.\ 12, this star falls to the red of the main sequence 
and below the turn-off point and therefore we also reject the possibility that NGC\,6866-7 is a 
blue--straggler.

\begin{figure}[htb]
\includegraphics{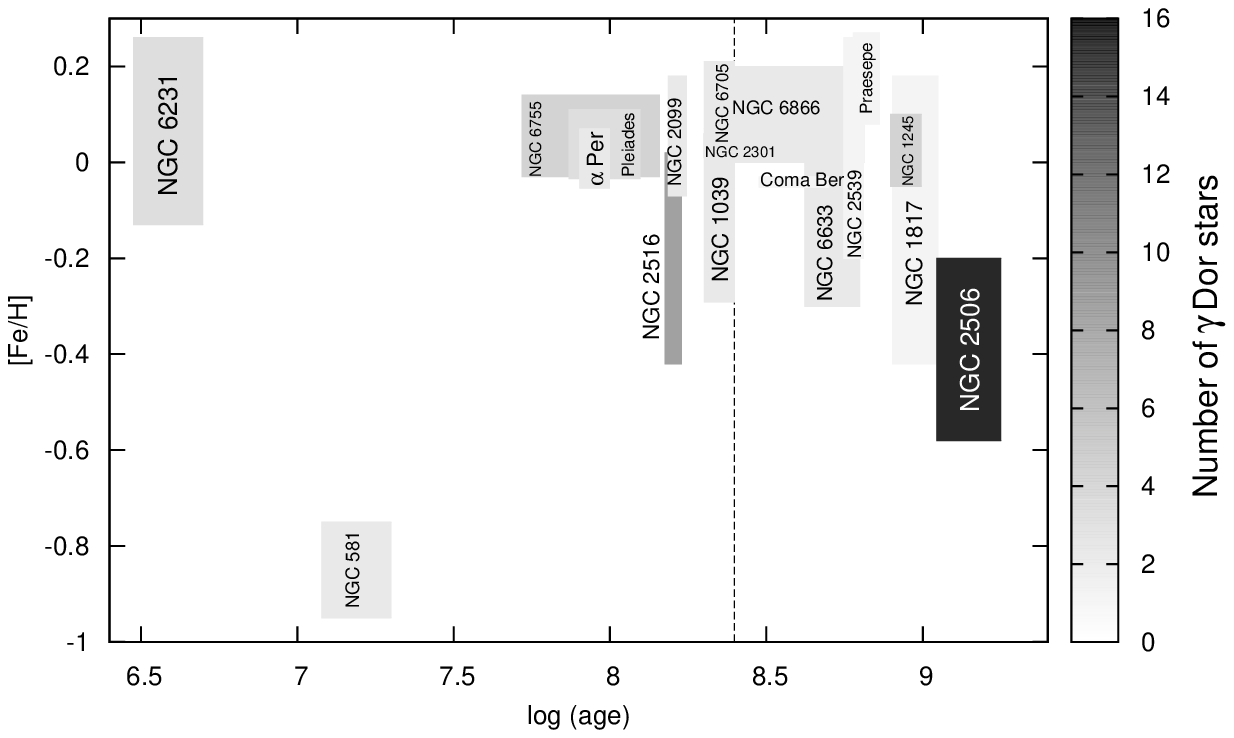}
\FigCap{The number of $\gamma$ Dor stars in open clusters coded with shades of gray. 
The size of each box represents the uncertainty of determination of the age
and $\rm [Fe/H]$ of the cluster. The vertical line at 250 Myr indicates a suspected 
upper limit of the age of an open cluster that can host $\gamma$ Dor stars (see Krisciunas and Patten 1999).
}
\end{figure}

In Fig.\ 12, there are, however, two other stars, NGC\,6866-1214 and TYC-2\,3162-00108-2 (which is the 
north--east faint component of NGC\,6866-5155), that fulfill all the requirements of bona fide 
blue--stragglers specified by Ahumada \& Lapasset (2007): their cluster membership probability is high:
79\,\%, 52\,\% or 90\,\% for NGC\,6866-1214, and 74\,\%, 62\,\% or 54\,\% for TYC-2\,3162-00108-2, as 
computed by Dias \etal (2002), Kharchenko \etal (2004) and in this paper, respectively, (although
Frinchaboy \etal (2008) give $P=0$\,\% for both stars), both fall into 
the area of the color-magnitude diagram where blue--stragglers are expected (cf.\ Fig.\ 12 of this paper 
and Figs.\ 1 and 2 of Ahumada \& Lapasset 2007), both are fainter than 2.5 mag limit above the cluster 
turnoff point, and fall close to the center of the cluster.

Therefore, we classify NGC\,6866-1214 and TYC-2\,3162-00108-2 as the first blue--stragglers discovered in 
the field of NGC\,6866. 

\section{Discussion}

Hosting one $\gamma$ Dor star and one $\gamma$ Dor candidate, NGC\,6866 joins the group of 20 galactic open 
clusters in which a total of 62 candidates for $\gamma$ Dor type pulsators have been discovered. We list these 
clusters in Table 4, where we give the number of $\gamma$ Dor candidates in the cluster (including the 
non-member $\gamma$ Dor stars, nm, the number of which, if known, is given below in brackets), and the range of the 
determinations of the age and metallicity of the cluster. The numbers in square brackets refer to the papers
listed at the bottom of the Table. In Fig.\ 13, we use the shades of gray to plot the number of $\gamma$ Dor 
candidates in the $\log (\rm {age})$ -- $\rm [Fe/H]$ plane. Below we discuss the properties of clusters that host $\gamma$ 
Dor candidates.

First, we see that, as already noticed by Eyer \etal (2002), there is no relationship between the age of an open cluster 
and the incidence of $\gamma$ Dor stars. The phenomenon of $\gamma$ Dor pulsators can occur in clusters as young 
as NGC\,6231 (3\,Myr, Baume \etal 1999) and as old as NGC\,2506 (1.8\,Gyr, Kim \etal 2001a). An upper limit
of the age of $\gamma$ Dor -- hosting open clusters, equal to 250 Myr according to Krisciunas \& Patten (1999), 
does not exist, because around 40\,\% of the clusters is older than that.

Second, $\gamma$ Dor stars seem to have little preference concerning the metallicity of the cluster; 
$\rm [Fe/H]$ of all but one cluster from Table 4 falls into the range from $-0.2$ to $+0.2$ dex to within its error bars.
Moreover, candidates for $\gamma$ Dor stars occur in clusters as metal-rich as Praesepe, $\rm [Fe/H] = +0.27$,
and as metal-poor as NGC\,581, $\rm [Fe/H] = -0.85$. We note also that although NGC\,2506 and NGC\,2516, i.e., 
the clusters in which the number of $\gamma$ Dor candidates is the highest, are the most metal-deficient ones, 
it needs to be confirmed how many of their $\gamma$ Dor candidates really pulsate and how many are cluster members.

The membership of $\gamma$ Dor candidates in a cluster is another issue that has not been studied sufficiently. 
For eight clusters only, NGC\,581, 1817, 2301, 2516, 6633, 6705, 6866, and 7086, the probability of cluster 
membership has been derived from the proper motions. For four clusters, NGC\,1817, 2516, 6705, and 6866, both 
proper motions and color-magnitude diagrams were used, leading to different results for some stars (see 
Arentoft \etal 2005). For NGC\,659, 2099, 2506, 2539, 6231, 6755, 7762, and the Pleiades, the cluster membership 
was determined only from the color-magnitude diagrams, for NGC\,1245, from the background star counts (Burke \etal 
2004), and for NGC\,1039, $\alpha$ Per, Praesepe, and Coma Ber, the probability of membership has not been computed.

The last but not least issue is the mechanism of the observed variability of these stars. All the 22 
monoperiodic $\gamma$ Dor candidates detected in the open clusters need to be confirmed as pulsators while the 
multiperiodic ones require further study to check 
which of the detected frequencies are due to pulsations and which to, e.g., Ell or $\alpha ^2$CVn type of variability.

Summarizing, we conclude that at the present stage it is not possible to find statistically significant 
relation between the age or metallicity of an open cluster and the number of $\gamma$ Dor stars therein.

\Acknow{This work was supported by MNiSW grant N203 014 31/2650 and the University of Wroc{\l}aw grant No 
2646/W/IA/06. We acknowledge A.\ Pigulski and G.\ Michalska for taking some observations used in this paper. 
We also thank Prof.\ M.\ Jerzykiewicz for his kind advice which helped us improve our paper. The authors 
made use of the WEBDA database, operated at the Institute for Astronomy of the University of Vienna, and 
of the NASA's Astrophysics Data System.}

\end{document}